\documentstyle[aaspptwo]{article}
\def\mg2{Mg$_2$}

\def\ha{\relax \ifmmode {\rm H}\alpha\else H$\alpha$\fi}
\def\hb{\relax \ifmmode {\rm H}\beta\else H$\beta$\fi}
\def\hi{\relax \ifmmode {\rm H\,{\sc i}}\else H\,{\sc i}\fi}
\def\hii{\relax \ifmmode {\rm H\,{\sc ii}}\else H\,{\sc ii}\fi}
\def\h2{\relax \ifmmode {\rm H}_2\else H$_2$\fi}
\def\lha{\relax \ifmmode L_{{\rm H}\alpha}\else $L_{{\rm H}\alpha}$\fi}
\def\shi{\relax \ifmmode \sigma_{{\rm HI}}\else $\sigma_{\rm HI}$\fi}   
\def\sh2{\relax \ifmmode \sigma_{{\rm H}_2}\else $\sigma_{{\rm H}_2}$\fi}
\def\kms{\relax \ifmmode {\,\rm km\,s}^{-1}\else \,km\,s$^{-1}$\fi}
\def\deg{\hbox{$^\circ$}}
\def\min{\hbox{$^\prime$}}
\def\sec{\hbox{$^{\prime\prime}$}}
\def\fdg{\hbox{$.\!\!^\circ$}}

\def\farcm{\hbox{$.\mkern-4mu^\prime$}}
\def\farcs{\hbox{$.\!\!^{\prime\prime}$}}
\def\degd#1.#2{ #1\fdg#2 }                 % degrees over decimal point     
\def\mind#1.#2{ #1\farcm#2 }               % minutes over decimal point
\def\secd#1.#2{ #1\farcs#2 }               % seconds over decimal point

\def\gtorder{\mathrel{\raise.3ex\hbox{$>$}\mkern-14mu
    \lower0.6ex\hbox{$\sim$}}}
\def\ltorder{\mathrel{\raise.3ex\hbox{$<$}\mkern-14mu
    \lower0.6ex\hbox{$\sim$}}}

\begin{document}

\title{A subarcsecond resolution near-infrared study of Seyfert and
`normal'  galaxies: II.~Morphology}

\author{Johan H. Knapen$^{1}$}
\author{Isaac Shlosman$^{2}$}
\author{Reynier F. Peletier$^{3,4}$}

\affil{$^{1}$Department of Physical Sciences,
University of Hertfordshire,
Hatfield, Herts AL10 9AB, UK,
E-mail: knapen@star.herts.ac.uk}

\affil{$^{2}$ Department of Physics \& Astronomy,
University of Kentucky,
Lexington, KY 40506-0055, USA,
E-mail: shlosman@pa.uky.edu}

\affil{$^{3}$Dept. of Physics, 
University of Durham, South Road,
Durham, DH1 3LE, UK}

\affil{$^{4}$School of Physics and Astronomy, 
University of Nottingham,
Nottingham, NG7 2RD, UK,
E-mail: Reynier.Peletier@Nottingham.ac.uk}

\journalid{Vol}{Journ. Date}
\articleid{start page}{end page}
\paperid{manuscript id}
\cpright{type}{year}
\ccc{code}
\lefthead{}
\righthead{}

\begin{abstract}

We present a detailed study of the bar fraction in the CfA sample of
Seyfert galaxies, and in a carefully selected control sample of
non-active galaxies, to investigate the relation between the presence of
bars and of nuclear activity.  To avoid the problems related to bar
classification in the RC3, e.g., subjectivity, low resolution and
contamination by dust, we have developed an objective bar classification
method, which we conservatively apply to our new sub-arcsecond
resolution near-infrared imaging data set (Peletier et al. 1999). We are
able to use stringent criteria based on radial profiles of ellipticity
and major axis position angle to determine the presence of a bar and its
axial ratio.  Concentrating on non-interacting galaxies in our sample
for which morphological information can be obtained, we find that
Seyfert hosts are barred more often (79\%$\pm$7.5\%) than the non-active
galaxies in our control sample (59\%$\pm$9\%), a result which is at the
$\sim2.5\sigma$ significance level.  The fraction of non-axisymmetric
hosts becomes even larger when interacting galaxies are taken into
account. We discuss the implications of this result for the fueling of
central activity by large-scale bars. This paper improves on previous
work by means of imaging at higher spatial resolution and by the use
of a set of stringent criteria for bar presence, and confirms that the
use of NIR is superior to optical imaging for detection of bars in disk
galaxies.  \end{abstract}


\section{Introduction}

The study of fueling processes in active galactic nuclei (AGNs) is an
issue important for our understanding of structure and evolution both of
central engines in AGNs and of their host galaxies. Although fuel is
plentiful in the disk, it needs to overcome the centrifugal barrier to
reach the innermost regions in disk and elliptical galaxies.
Large-scale non-axisymmetries, such as galactic bars, are thought to be
related to starburst activity within the central kpc, which
preferentially occurs in barred hosts (e.g., Heckman 1980; Balzano 1983;
Devereux 1987; Kennicutt 1994). In  a number of early optical
surveys, the fueling of Seyfert (Sy) activity in disk galaxies was
linked to  non-axisymmetric distortions of galactic gravitational
potentials by large-scale stellar bars and tidal interactions  (Adams
1977; Heckman 1978; Simkin, Su \& Schwarz 1980; Dahari 1984). This was
supported by a more superficial argument that gravitational torques are
able to remove the excess angular momentum from gas, which falls
inwards, giving rise to different types of activity at the center (see
reviews by Shlosman 1992; Sellwood \& Wilkinson 1993; Phinney
1994). However, a criticism was leveled on observational results
because the control samples were not matched to the Sy sample in
properties like morphological distribution (Balick \& Heckman 1982;
Fuentes-Williams \& Stocke 1988; Shlosman, Begelman \& Frank 1990).
More recent studies by Moles, M\'arquez \& P\'erez (1995) and Ho,
Filippenko \& Sargent (1997) conclude that the fraction of barred
galaxies is equal among AGN hosts and the general population. However,
these studies, like the ones in the past, continue to rely on (a) the
morphological classification from optical catalogs (e.g., RC3; de
Vaucouleurs et al. 1991), or (b) the use of control samples that are not
matched to the ``active'' samples, e.g., the general population of
spirals in the RC3.

A combination of dust obscuration, stellar populations and inadequate
spatial resolution can hide even a strong bar in the optical (Thronson
et al. 1989, Block \& Wainscoat 1991; Spillar et al. 1992). High-quality
near-infrared (NIR) imaging is much more reliable in determining the
overall mass distribution in galaxies. Mulchaey \& Regan (1997) much
improved the observational basis of this area of study, using NIR
imaging of matched Sy and control samples of galaxies. Although their
use of NIR imaging results in slightly higher bar fractions than, e.g.,
Ho et al. (1997), Mulchaey \& Regan do not find evidence for a
significant excess of bars in Sy galaxies. Recent work by Regan \&
Mulchaey (1999) involving HST color maps to search for the dust lane
signatures of small-scale bars is potentially promising, but employs a
very small and statistically insignificant sample. It is also unclear
whether nuclear bars have similar properties (e.g., dust lanes) to
large-scale stellar bars, and, hence, whether the criterion used is
reliable.

The present paper aims to improve on past work through a combination of
the following factors:

\begin{enumerate}

\item use of NIR observations of a consistently high (sub-arcsecond)
spatial resolution

\item use of a carefully selected control sample of non-active galaxies

\item use of objective and stringent criteria for assigning bars, and
for determining bar axial ratio

\item publication of our complete set of images and profiles 
(Peletier et al. 1999, hereafter Paper~I), to allow intercomparison with
samples of other workers.

\end{enumerate}

This paper is the second in a series analyzing the circumnuclear
morphology and colors of a sample of Sy galaxies. Paper~I describes the
sample, the observations, data reduction and photometric calibration,
and an analysis of the circumnuclear regions and host galaxy disks
through color index maps. Paper~III (Shlosman, Peletier \& Knapen 1999)
focuses on the distribution of bar axial ratios in Sy and non-active
galaxies. In the present paper (Paper~II) we compare the morphology of
the Sy sample with that of a control sample of non-active galaxies,
determining the fraction of non-axisymmetric structures in AGN hosts.

In Sect.~2, we analyze the statistics of bars as determined from the
RC3, and use that analysis in Sect.~3 to describe our Sy and control
samples. Sect.~4 reviews the images and profiles of sample objects, as
published in detail in Paper~I, and describes our criteria for
recognizing a bar in a given galaxy, and for determining the bar's
ellipticity. We present our results on bar fractions in Sect.~5, and
compare them with other published work in Sect.~6. Finally, our results
are discussed in terms of scenarios describing fueling of central
activity by large-scale stellar bars in Sect.~7.

\section{Bar statistics from the RC3}

Before studying in detail the occurrence of bars in galaxies as
determined from high-resolution NIR images, we consider statistics of
bars as obtained from the RC3 catalogue. Although NIR imaging has a
better sensitivity to barred potentials, which we exploit in the rest of
the paper, use of the RC3 catalogue has the advantage of sample
size. Classifications for literally thousands of galaxies can be used to
obtain indications on how bar fractions in galaxy populations vary as a
function of various parameters.

Fig.~1 shows how many galaxies are classified as `barred' (type X or B)
in the RC3 by displaying the fraction of barred galaxies as a function
of morphological type (Fig.~1a), of ellipticity $\epsilon = 1-b/a$ 
(equivalent to inclination,Fig.~1b),  and of  recessional velocity
(or redshift, Fig.~1c). All parameters are directly obtained from the
RC3.

\begin{figure}
\mbox{\epsfysize=16cm  \epsfbox{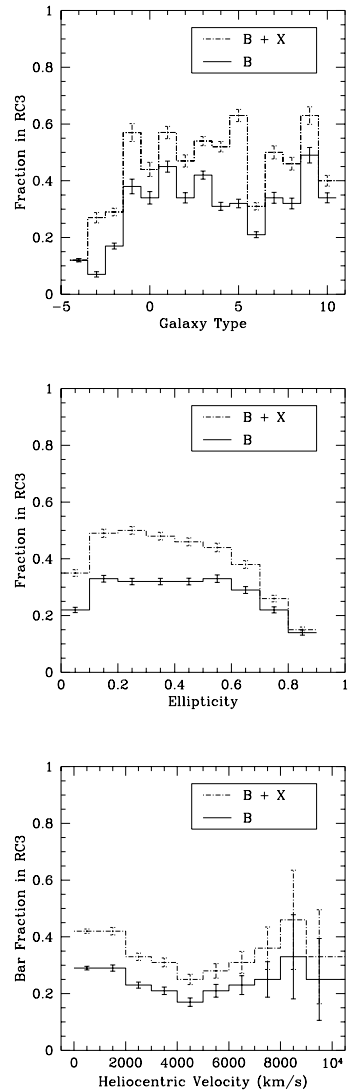}}
\caption{
Bar fractions as determined from the RC3 morphological classification as
a function of {\it (a)} morphological type, {\it (b)} galaxy ellipticity
$(1-b/a)$, and {\it (c)} heliocentric velocity.}
\end{figure}

Fig.~1 confirms an overall bar fraction of $\sim 50-60$\% (e.g.,
Sellwood \& Wilkinson 1993), but also shows that the bar fraction hardly
varies with (1) morphological type for lenticular and spiral galaxies,
(2) observed (i.e., not deprojected) ellipticity $\epsilon$, within the
range of $0.1-0.4$, and even 0.6 for the `B' galaxies, and (3)
heliocentric velocity or $z$. Due to the small numbers of galaxies at
$v > 10^4\kms$, the results on the distribution with  recessional
velocity are not conclusive, but do not show large systematic
effects. As expected, the bar fraction is substantially lower for
galaxies of extremely early type, and for highly inclined galaxies
($\epsilon > 0.6$), where a bar will be much harder to recognize or
classify.

Using the morphological classification of the complete set of galaxies
in the RC3, we can determine the RC3 bar fractions in our Sy sample (see
Section~3 for a description of the sample selection), and of a synthetic
sub-sample of normal galaxies. The latter sub-sample was constructed
from the RC3, with a distribution in terms of morphological type,
ellipticity, and absolute magnitude similar to that of the Sy sample.

Since we have not encountered significant systematic effects in the RC3
classification within the parameter space under consideration here (see
above), we will not at this stage correct the RC3 morphologies for
systematic effects. We match the RC3 sub-sample to the Sy sample in the
following way. For each CfA Sy galaxy of type t$_i$, ellipticity
$\epsilon_i$ and absolute magnitude M$_i$, we select the galaxies in the
RC3 with these parameters, and determine the barred fraction ($f_{\rm
t_i,\epsilon_i,\rm M_i}$). The fraction of barred and mixed type
galaxies for the synthetic RC3 sub-sample thus constructed is given by $
f_{\rm{B,Sy}}={\Sigma_{i=1}^{n}f_{\rm t_i,\epsilon_i,\rm M_i}/{n}}$,
where $n$ is the number of Sy's.

Using the criteria outlined in more detail in Sect.~4, we excluded all
galaxies for which one can not perform a reliable bar classification.
The CfA sample thus reduces from 48 to 29 Sy's. Of these 29, 8 objects
(28 $\pm$ 8\%) have been classified in the RC3 as type B, and 7 (24
$\pm$ 8\%) of type X (total B+X 52 $\pm$ 9\%). For the corresponding
synthetic RC3 sub-sample the fraction of B-type galaxies is 40 $\pm$
2\%, and of mixed type 22 $\pm$ 2\% (total B$+$X 62 $\pm$ 2\%). 
Although there is no  clear correspondence for individual galaxies
between the RC3 morphological types and the bar axial ratios (Paper III;
also R.J. Buta, private communication),  these numbers do indicate a
tendency for B-type bars to be underabundant in the Sy sample, a finding
which is upheld by studying other samples  (Paper III).

We conclude that {\it based solely on the RC3} optical classification,
there is a slight, and not necessarily significant, deficiency of barred
galaxies among the Sy sample as compared to a synthetic control
sub-sample with the same properties. This cannot be caused by any of the
selection effects in morphological type, ellipticity, or absolute
magnitude, as outlined above, because the two samples were explicitly
matched using those criteria.

\section{Observations and sample selection}

We use the set of NIR imaging observations of the (CfA) sample of Huchra
\& Burg (1992) for the Sy's, as described in Paper~I. We only summarize
the basic details here. We observed the complete CfA sample in the NIR
$J, H$ and $K$ bands, apart from 3C~273 which was not included because
of its high redshift (0.16 vs. 0.07 for the second highest).
Observations were made under good seeing conditions (spatial resolution
of $<1''$, best seeing $\secd 0.55$, with median over all images $\secd
0.7$). The galaxy exposures were interleaved with sky exposures, and
standard stars were observed for the photometric calibration.

The data of our sample of 48 Sy galaxies are compared with $K$-band
images of the same quality of a control sample of 34 galaxies, selected
from the RC3 to mimic the Sy sample in terms of morphological type and
ellipticity. Full details of the observations and data reduction can be
found in Paper~I, where we also publish all the images and derived
surface brightness, color, ellipticity and position angle profiles for
both Sy and non-Sy samples. We made sure none of our control
galaxies is active by cross-checking with different sources of AGN
classification. In Fig.~2 we provide the distribution of both
samples as a function of morphological type,  galaxy ellipticity and
redshift. One important difference between the two samples, obvious in
these figures, is that a significant fraction of the Sy galaxies does
not have a well-defined type in the RC3. For such galaxies, as well as
for those at high inclination, or large $z$, determination of the
presence and axial ratios of bars will be difficult or impossible, and we
thus proceeded by removing those galaxies from our sample. 

\begin{figure}
\mbox{\epsfxsize=6.5cm  \epsfbox{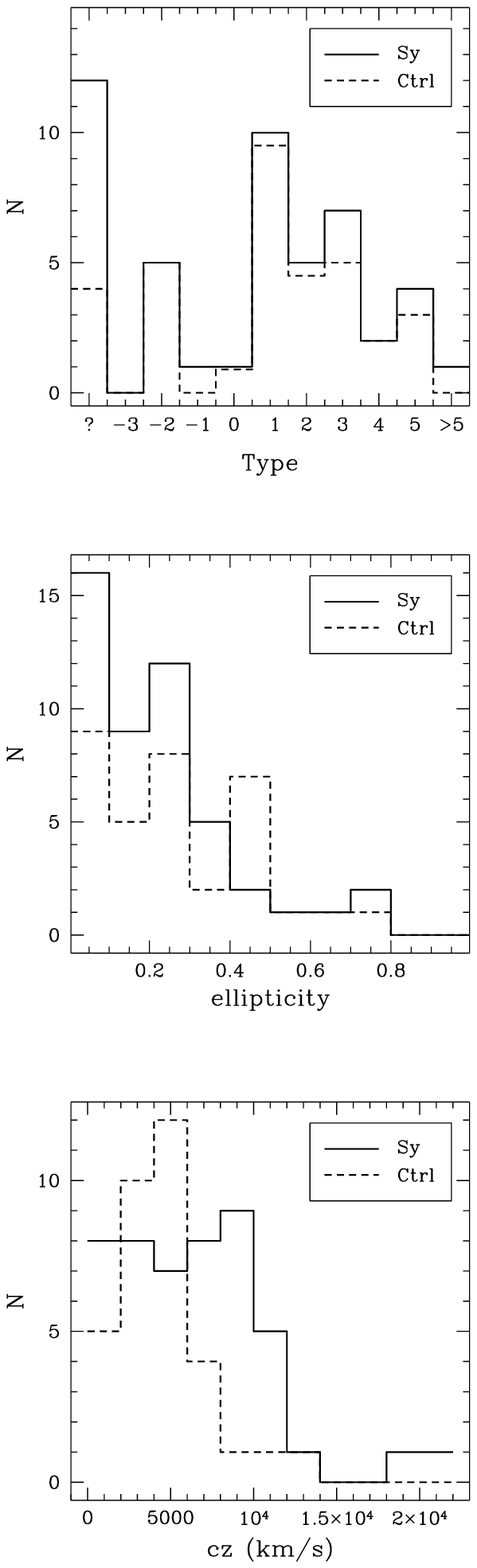}}
\caption{
Distribution of Sy and control sample before removing the
galaxies for which no reliable bar classification could be made.}
\end{figure}

We  exclude from our analysis those galaxies that are

\begin{itemize}

\item too small to study their morphology, as determined from their
radius (in arcsec) at a surface brightness level of 19 $H$-mag\,
arcsec$^{-1}$: if $\log r_{H, 19}<0.8$, the galaxy is disqualified. Such
galaxies have been marked with code~1 in Tables~1 and 2.

\item strongly interacting, as evident either by a severely distorted
morphology, or by the presence of a companion within 1\min\ (code~2).

\item highly inclined, with $\epsilon>0.5$ (see Fig.~1b) (code~3).

\end{itemize}

We chose the limits in such a way as to disqualify those galaxies for
which the information on the isophote shapes as derived from the imaging
would not lead to reliable statements on the presence of a bar. The
final statistical results do not critically depend on fine-tuning these
numerical limits.

In Fig.~3 we show the distributions for the reduced sample, which we use
in the statistical analysis (Sect.~5). Although there are still some
minor differences between the reduced Sy and non-Sy samples in terms of
morphological type, ellipticity and redshift, such differences are
small. In any case, the remaining differences are unlikely to affect our
statistical results on bar fractions, because we have shown in Fig.~1
that there are no systematic trends of bar fraction as a function of
these parameters, over the range spanned in type, ellipticity and $z$.

\begin{figure}
\mbox{\epsfxsize=6.5cm  \epsfbox{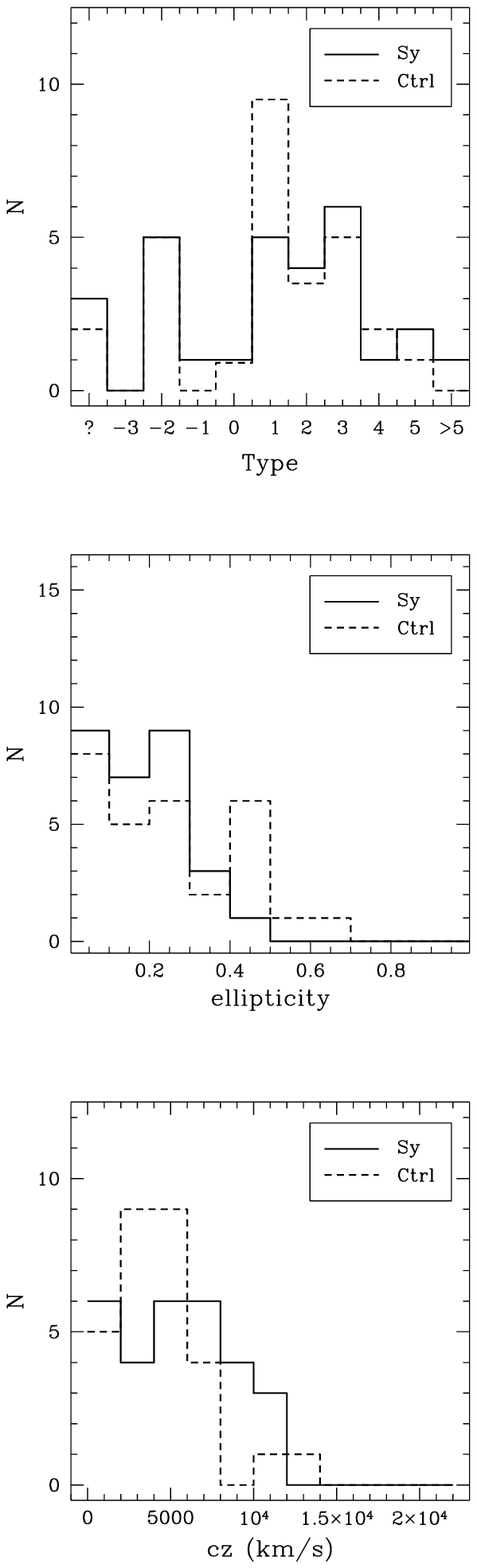}}
\caption{
Distribution of Sy and control sample after removing the
galaxies for which no reliable bar classification could be made.}
\end{figure}

\section{Bar criteria and determination of axial ratios}

Our main criterion for the presence of a bar is a significant rise in
the radial ellipticity profile of a galaxy (Paper~I), followed by a
significant fall, over a range in radius where the position angle of the
major axis is roughly constant. The amplitude of the ellipticity
variation must be at least 0.1.

We also classify galaxies as barred if their radial major axis position
angle profile (Paper~I) shows a change of more than 75\deg, accompanied
by ellipticity values above roughly 0.1. Such NIR isophote twists are
characteristic of the inner disk resonance region where nuclear rings
are found (Shaw et al. 1993; Knapen et al. 1995; Heller \& Shlosman
1996). Furthermore, large isophote twists ($>$ 50$^{\rm o}$) are very
uncommon in elliptical galaxies, and hence it is very unlikely that they
are related to the possible triaxiality of the central bulges in disk
galaxies (Peletier et al. 1990).

We primarily used our NIR imaging, but since most galaxies are larger
than our field of view, we extended the radial range of the ellipticity
and position angle profiles by using images from the digitized sky
survey (see Paper~I). This allows the detection of large-scale bars, for
which our NIR data alone would not be sufficient to confirm the decrease
in ellipticity beyond the range of the bar.

In a number of cases, we suspect the presence of a bar in a galaxy, but
the results obtained from our data do not satisfy our stringent bar
criteria. We have thus counted these galaxies as non-barred in our
statistics, but data of higher quality may well lead to its
re-classification as barred galaxies. This group includes as most
prominent members Mrk~461 and UGC~8621 in the Sy sample, and NGC~3455
and NGC~4966 in the control sample.

We determined axial ratios  of the bars from the ellipticity profiles
using the criterion put forward by Martin (1995), where $\epsilon_{\rm
b}=10(1-b/a)$, thus `no bar' means $\epsilon_{\rm b}=0$, and a very
`strong' bar $\epsilon_{\rm b}=8$. $\epsilon_{\rm b}$ is read off from
the profile as the maximum $\epsilon$ at the bar position, and is
subsequently corrected for the inclination of the galaxy. The bar
ellipticity is used here in a narrow  sense, namely to define the
axial ratios. Realistic bars of course can differ substantially from
ideal elliptical shapes.

Tables~1 and 2 contain information on the  presence of bars in our
sample galaxies, and on the maximum axial ratios of those bars. An entry
in the ``Bar `strength' '' column indicates the presence of a bar, a
blank in that column accompanied by a blank in the last column (8,
``Code'') means the galaxy is non-barred in our imaging study.

\section{Results}

\subsection{CfA Sy sample}

After disqualifying galaxies due to their small angular size,
interaction, or high inclination, our Sy sample is reduced from 48
galaxies to 29. Of these 29 galaxies, 20 have a bar as determined from
the ellipticity profile, and for all of these we can determine the
bar axial ratio. Another 3 galaxies are barred as determined from their
position angle twist, but for these the bar axial ratios cannot be
determined. We conclude that 23 out of 29 galaxies are barred, or
79\%$\pm$7.5\% ($1\sigma$ Poisson error).

Mrk~270 has a small bar at $r<2\sec$, but since this lies at the edge of
our resolution limit (about three times the seeing value) we have taken
the convincing position angle twist at larger radii as prime evidence
for the presence of a bar.

\subsection{Control sample}

Following the same approach as for the Sy galaxies, we first exclude 5
of our 34 galaxies due to interactions or high inclination. Of the
remaining 29, 17 are found to be barred, or 59\%$\pm$9\%. We could
determine the bar axial ratios for all but two  of these galaxies.

\section{Comparison with published results}

As outlined in the introduction, the methodology used in the present
paper is very different from that used in most related published
studies, with the exception of Mulchaey \& Regan (1997). All other
authors used morphological classifications as taken at face value from
one of the main galaxy catalogues. Furthermore, in many studies, an
active galaxy population is compared with a control population which is
not matched in detail.

Ho et al. (1997) use the results from their spectrocopic survey of a
large magnitude-limited sample of galaxies to study the relations
between bar fraction and nuclear activity. Information on activity in
their sample galaxies is obtained from this spectral survey, whereas
morphological information on the host galaxy is taken from the RC3. Ho
et al.  consider an ``AGN'' subsample, which includes LINER/\hii\
transition objects, LINERs and Sy's, and compare its bar fraction with
that of the complete sample. They find that the bar fraction in the AGN
sample is slightly lower, but not significantly so (57$\pm$4\% for the
AGN sample vs. 61$\pm$4\%).

Mulchaey \& Regan (1997) invoked a similar (but not identical in
details) approach to the one described by us here, namely analyzing NIR
images of matching samples of Sy and non-active galaxies. To determine
the presence and axial ratios of the bars, they use in the first instance
radial ellipticity and position angle profiles (as published by
Mulchaey, Regan \& Kundu 1997), but, unlike in the present study, add a
few cases where the bar is ``visible'' on the image, but not in the
profiles. Mulchaey \& Regan (1997) conclude that the galaxies in both
their Sy and control samples are barred in just over 70\% of all
cases. Based on the size of their samples, the errors in these numbers
should be around 8\%.  Considering these uncertainties, their result is
not in contradiction with results of this study.

The main improvement in our work is the use of images at consistently
higher resolution than Mulchaey \& Regan (median seeing of our data
\secd 0.7, vs.  \secd 1.0, with variations between \secd 0.8 and \secd
1.8, for Mulchaey \& Regan 1997).  Some other differences between the
study of Mulchaey \& Regan (1997) and  ours, that might influence
the resulting bar fractions, can be identified. Firstly, Mulchaey \&
Regan use a subset of the incomplete sample of Maiolino et al. (1995),
essentially all known Sy's in the RSA (Sandage \& Tammann 1981). The CfA
sample,  on the other hand, is considered to be an optically complete
set of Sy's in a region of the sky.  Secondly, Mulchaey \& Regan match
their active and control samples using, among other parameters, the
absolute blue magnitude, whereas we chose not to. The advantage  of
Mulchaey \& Regans approach is that the two samples would be better
matched in distance, but the uncertainty about the fraction of total
blue light attributed to the galaxy by the AGN, and the, potentially
related, fact that Sy host galaxies tend to be more luminous than
average spirals (Mulchaey \& Regan 1997), in our opinion offset this
advantage. Thirdly, whereas Mulchaey \& Regan, in a few cases, update
their bar classification on the basis of a visual inspection, even if
the ellipticity and position angle profiles do not convincingly show the
presence of a bar, we have stubbornly maintained our stringent criteria
for bar presence based upon the behavior of the radial profiles. This
may result in a lower bar fraction than that found by Mulchaey \& Regan,
as is in fact the case in the control sample. Our higher bar fraction in
the Sy sample, then, is further proof of the power of a modest increase
in spatial resolution in the NIR imaging.

We adjust the classification given by McLeod \& Rieke (1995) in two
cases: NGC~3982, classified as  .SXR3*. (or SAB(r)b) in the RC3 (but
not barred according to McLeod \& Rieke) does in fact have a bar inside
a region of well-defined spiral arms; and Mrk~817 has a nice but
angularly small bar.

We have cross-checked our results for the presence of bars in the Sy
host galaxies with those presented by Malkan et al. (1998). The
following galaxies in our CfA sample were classified as barred by Malkan
et al.: Mrk~993 (disqualified by us due to its high inclination),
Mrk~766, NGC~5674, Mrk~817, NGC~5940, NGC~6104 (disqualified by us for
interaction), UGC~12138, Mrk~533 (disqualified by us for interaction),
and NGC~7682. Apart from the  three galaxies we disqualified, all
these galaxies are confirmed to be barred in our study.

An extensive comparison of bar axial ratios as determined by us and taken
from the literature, statistically and in individual cases, is presented
in Paper~III.

\section{Concluding remarks: large-scale stellar bars and fueling of AGNs}

Based on our new sub-arcsec resolution NIR imaging survey of the CfA
sample of Sy's and of a control sample of normal galaxies, and excluding
objects for which reliable morphological information cannot be obtained,
we find that Sy hosts are barred more often than normal galaxies
(79\%$\pm$7.5\% barred for the Sy's, vs. 59\%$\pm$9\% for the control
sample). This result stands at the 2.5$\sigma$ level. 

Taking the bar classification from the RC3, we find that Sy's are barred
just as often as normal galaxies, in general agreement with previous
work (e.g., Simkin et al. 1980; Moles et al. 1995; Ho et al. 1997), 
also based upon the RC3. The classification in the RC3 was done by eye on the
basis of photographic plates taken in the optical and at relatively poor
angular resolution, and although performed by experts, the
classification remains difficult to reproducible for individual
galaxies. This is illustrated, for example, by Sellwood \& Wilkinson
(1993), who compare bar fractions based on several catalogues, and by
the present work (see Tables~1 and 2).

In contrast, in our new study we observe  the morphology of active and
non-active galaxies in the NIR, unhampered by dust, and at a much higher
resolution than previous work, including the other NIR studies in
this field, by McLeod \& Rieke (1995) and Mulchaey \& Regan (1997). Our
result, at a significance level of $\sim2.5\sigma$, suggests that there
is an underlying morphological difference between Sy and non-Sy
galaxies. Since the large error margins are a direct result of the
sample size, it is now imperative to check the validity of our
observational conclusion with larger samples.

A few comments follow from the above statistics. Firstly, these results
emphasize the prevalence of barred morphologies in disk galaxies in
general, and in active galaxies in particular. It is known from
extensive theoretical studies that non-axisymmetric potentials induce
radial gas flows and elevate dramatically the rate of star formation
within the central kpc of barred galaxies. A straightforward
extrapolation towards much smaller spatial scales characterizing the
non-stellar AGN-type activity is not warranted however. The absence of
AGNs in many barred galaxies can be interpreted in terms of additional
factors, besides the large-scale stellar bars, which are required to
trigger the nuclear nonstellar activity -- a point emphasized already by
Shlosman, Frank \& Begelman (1989). These additional factors should
include the availability of fuel, efficiency of star formation and
global self-gravitating effects in the circumnuclear gas, underlining
the increasingly important role the gas plays in galactic dynamics at
progressively smaller radii (Shlosman 1996).

To summarize, our new NIR data set improves on earlier published surveys
of the CfA Sy sample in the following aspects: (1) complete coverage of
the sample, (2) observations in the NIR $J$, $H$ and $K$ bands of all
objects, (3) improved spatial resolution: $<1''$ for all our images,
best seeing $\secd 0.55$, with median over all images $\secd 0.7$, and
(4) observations of a matched control sample in the $K$ band. In
addition, there is no reliance on  optically selected galaxy
catalogues, and the morphology  of both the CfA and the control
samples  galaxies was critically revised. {\it At the resolution
level of RC3 optical catalog}, there is an agreement between our results
and those of past and recent surveys (e.g., Simkin et al. 1980; Moles et
al. 1995; Ho et al. 1997). It is the high NIR resolution used here that
led us to our main conclusion.

ACKNOWLEDGEMENTS.  We thank Ron Buta, Fran\c{c}oise Combes and Bill Keel
for helpful communications. We made use of the NASA/IPAC Extragalactic
Database (NED) which is operated by JPL, CalTech, under contract with
NASA. IS is supported in part by NASA grants NAGW-3841,
WKU-522762-98-06, and HST AR-07982.01-96A.

\vfill\eject

\vfill\eject
\onecolumn

\begin{table}[h]
\small
\begin{center}
\begin{tabular}{llcccccl}
\\
\\
\hline
\hline
Galaxy & v$_{\rm hel}$ & Sy & T & Class. 
 & Bar `strength' & Bar `strength' & Code \\
 ~ & ~ & ~ & ~ & ~ & Projected & Deprojected & ~ \\
\hline
Mrk 334   &6582& 1.8 &  N   &  .P.....   &   & & 2 \\
Mrk 335   &7688& 1 &  N   &  .P.....     &   & & 1 \\
UGC 524   &10763& 1 &  3.0 &  PSBS3..     & 6 & 6 & \\
I Zw 1    &18330& 1 &  N   &  .S?....     &   & & 1 \\
Mrk 993   &4658& 2 &  1.0   &  .S..1..   &   & & 3 \\
Mrk 573   &5174& 2 &  -1.0&  RLXT+*.     & ND & ND & 4 \\
UGC 1395  &5208& 1.9 &  3.0 &  .SAT3..   & 6 & 5 & \\
Mrk 590   &7910& 1.2 &  1.0 &  .SAS1*.   & ND & ND & 4 \\
NGC 1068  &1136& 2 &  3.0 &  RSAT3..     & 5 & 4 & \\
NGC 1144  &8648& 2 &  N   &  .RING.B     &   &  & 2 \\
Mrk 1243   &10602& 1 &  1.0 &  .S..1..     &  &  & 1\\ % = NGC 3080
NGC 3227  &1157& 1.5 &  1.0 &  .SXS1P.   &   &  & 2,3 \\
NGC 3362  &8318& 2 &  5.0 &  .SX.5..     & 4 & 2 & \\
UGC 6100  &8778& 2 &  1.0 &  .S..1?.     &   &  & \\
NGC 3516  &2649& 1.5 &  -2.0&  RLBS0*.   & 3 &  4 & \\
Mrk 744   &2674& 1.8 &  1.0 &  .SXT1P.   &   &  & 2,3\\
NGC 3982  &1109& 2 &  3.0 &  .SXR3*.     & 4 & 3 & \\
NGC 4051  &725& 1 &  4.0 &  .SXT4..     & 7 & 6 & \\
NGC 4151  &995& 1.5 &  2.0 &  PSXT2*.   & 5 & 3 & \\
NGC 4235  &2410& 1 &  1.0 &  .SAS1./     &   &  & 3 \\
Mrk 766   &3876& 1.5 &  1.0 &  PSBS1*.   & 6 & 6 & \\
Mrk 205   &21239& 1 &  N   &  .P.....     &   &  & 1\\
NGC 4388  &2524& 2 &  3.0 &  .SAS3*/     &   &  & 3\\
NGC 4395  &319& 1.8 &  9.0 &  .SAS9*.   & 7 & 7 & \\
Mrk 231   &12642& 1 &  5.0 &  .SAT5\$P    &   & & 2 \\
NGC 5033  &875& 1.9 &  5.0 &  .SAS5..   &   &  & 3\\
Mrk 789   &9476& 1 &  N   &  .P.....     &   &  & 1,2\\
UGC 8621  &6023& 1.8 &  N   &  .S?....   &   &  & \\
NGC 5252  &6926& 1.9 &  -2.0&  .L.....   & 6 & 4 & \\
Mrk 266   &8353& 2 &  N   &  .P.....     &   &  & 2\\
Mrk 270   &2700& 2 &  -2.0&  .L...?.     & ND & ND & 4\\
NGC 5273  &1089& 1.9 &  -2.0&  .LAS0..   &   &  & \\
Mrk 461   &4856& 2 &  N   &  .S.....     &   &  & \\
NGC 5347  &2335& 2 &  2.0 &  PSBT2..     & 7 & 6 & \\
Mrk 279   &8814& 1 &  -2.0&  .L.....     & 4 & 2 & \\
NGC 5548  &5149& 1.5 &  0.0 &  PSAS0..   &   &  & \\
NGC 5674  &7474& 1.9 &  5.0 &  .SX.5..   & 6 & 6 & \\
Mrk 817   &9430& 1.5 &  N   &  .S?....   & 5 & 5 & \\
Mrk 686   &4225& 2 &  3.0 &  .SB.3..     & 5 & 4 & \\
Mrk 841   &10852& 1 &  N   &  .......     &   &  & 1 \\
NGC 5929  &2492& 2 &  2.0 &  .S..2*P     &   &  & 2\\
NGC 5940  &10115& 1 &  2.0 &  .SB.2..     & 5 & 5 & \\
NGC 6104  &8382& 1.5 &  N   &  .S?....   &   &  & 2 \\
UGC 12138 &7375& 1.8 &  1.0 &  .SB.1..   & 6 & 5 & \\
NGC 7469  &4892& 1 &  1.0 &  PSXT1..     & 4  & 2 & \\
Mrk 530   &8851& 1.5 &  3.0 &  .SAT3*P   &   &  & \\
Mrk 533  &8713& 2 &  4.0 &  .SAR4P.     &   &  & 2\\ % = NGC 7674
NGC 7682  &5107& 2 &  2.0 &  .SBR2..     & 6 & 6 & \\
\hline
\end{tabular}
\end{center}
\hfil\break

\caption{Galaxy names (col.  1),  heliocentric radial velocity
(col. 2) and Sy classification (col. 3) from NASA's Extragalactic
Database (NED).  Numerical and morphological type (col.  4 \& 5) from
RC3.  Column 6 shows the bar strength (projected) as determined from the
ellipticity profiles (see text).  Code ND in this column means galaxy is
barred but no bar strength could be determined.  The deprojected bar
strength is given in column 7. Codes in col. 8 indicate whether galaxies
belong to any of the following groups: {\bf 1:} too small to study
morphology ($\log r_{H, 19}<0.8$ with $r$ in \sec) {\bf 2:} Strongly
interacting - severely distorted or companion within 1$'$; {\bf 3:}
Edge-on ($\epsilon$ $>$ 0.5); {\bf 4:} PA twist $> 75^{\rm o}$. Codes 1,
2 or 3 disqualify from further statistical analysis, code 4 indicates
presence of bar.}

\end{table}

\begin{table}[h]
\small
\begin{center}
\begin{tabular}{llcrccl}
\\
\\
\hline
\hline
Galaxy & v$_{\rm hel}$ & T &  Class.&   Bar `strength' & Bar `strength' & 
Code \\
 ~ & ~ & ~ & ~ & Projected & Deprojected & ~ \\
\hline
NGC 1093 & 6282& 2.0 & .SX.2?. &  7 &  6 & \\
UGC 3247 & 3371& N & .S?....   &  5 &  5 & \\
UGC 3407 & 3604& 1.0 & .S..1.. &  4 &  4 & \\
UGC 3463 & 2693& 4.0 & .SXS4.. &    &  & \\
UGC 3536 & 4689& -3.0 & .L.... &  6 &  2 & \\
UGC 3576 & 5966& 3.0 & .SBS3.. &  6 &  3 & \\
UGC 3592 & 13160& 1.0 & RSBS1.. &  7 &  8 & \\
UGC 3789 & 3325& 2.0 & RSAR2.. &  6 &  6 & \\
UGC 3850 & 4709& 1.0 & PSXS1.. &  5 &  5 & \\
NGC 2347 & 4421& 3.0 & PSAR3*. &    &  & \\
NGC 2365 & 2278& 1.0 & .SX.1.. &    &  & \\
NGC 2431 & 5679& 1.0 & PSBS1*. &  ND & ND & 4\\
NGC 2460 & 1442& 1.0 & .SAS1.. &    &  & \\
NGC 2487 & 4841& 3.0 & .SB.3.. &  6 &  5 & \\
NGC 2599 & 4741& 1.0 & .SA.1.. &    &  & \\
NGC 2855 & 1910& 0.0 & RSAT0.. &  3 &  2 & \\
NGC 3066 & 2049& 4.0 & PSXS4P. &  5 &  5 & \\
NGC 3188 & 7769& 2.0 & RSBR2.. &  6 &  6 & \\
NGC 3455 & 1102& 3.0 & PSXT3.. &    &  & \\
NGC 4146 & 6520& 1.5 & RSXS2.. &  5 &  5 & \\
NGC 4369 & 1045& 1.0 & RSAT1.. &  7 &  7 & \\
NGC 4956 & 4750& -2.0 & .L.....&    &  & \\
NGC 4966 & 7036& N & .S.....   &    &  & \\
NGC 5434 & 5638& 5.0 & .SA.5.. &   &  & 2,4\\
NGC 5534 & 2633& 1.7 & PSXS2P* &    &  & 2 \\
NGC 5832 & 447& 3.0 & .SBT3\$.&    &  & \\
NGC 5869 & 2087& -2.0 & .L..0*.&  ND  & ND & 4 \\
UGC 9965 & 4528& 5.0 & .SAT5.. &  5 & 4 &\\
NGC 5992 & 9518& N & .S.....   &  & & 2\\
NGC 6085 & 10195& 1.0 & .S..1.. &    &  & \\
NGC 6278 & 2790& -2.0 & .L.....&    & & \\
NGC 6504 & 4788& N & .S.....   &    & & 3\\
NGC 6635 & 5038& -2.0 & .L...P*&    & & \\
NGC 6922 & 5665& 5.3 & .SAT5P* &   & & 2\\
\hline
\end{tabular}
\end{center}
\caption{As Table 1, now for control sample}
\end{table}


\begin{references}

\reference{} Adams, T.F. 1977, ApJS, 33, 19
\reference{} Balick, B. \& Heckman, T.M. 1982, ARAA, 20, 431
\reference{} Balzano, V.A. 1983, ApJ, 268, 602 
%\reference{} Berentzen, I., Heller, C.H. \& Shlosman, I. 1997, MNRAS,
%   300, 49
\reference{} Block, D. \& Wainscoat, R.J. 1991, Nature, 353, 48
\reference{} Dahari, O. 1984, Ph.D. Thesis, University of California at 
   Santa Cruz
\reference{} de Vaucouleurs, G., de Vaucouleurs, A., Corwin, H.G., Buta,
   R.J., Paturel, G., Fouqu\'e, P. 1991, 3rd Reference Catalogue of Bright 
   Galaxies (RC3), Springer, New York
\reference{} Devereux, N.A. 1987, ApJ, 323, 91
\reference{} Fuentes-Williams, T. \& Stocke, J.T. 1988, AJ, 96, 1235
\reference{} Heckman, T. 1978, PASP, 90, 241
\reference{} Heckman, T. 1980, A\&A 88, 365
\reference{} Heller, C.H. \& Shlosman, I. 1996, ApJ, 471, 143
\reference{} Ho. L.C., Filippenko, A.V. \& Sargent, W.L.W. 1997, ApJ, 487,
   591
\reference{} Huchra, J.P., \& Burg, R. 1992, ApJ, 393, 90
\reference{} Kennicutt, R.C. 1994, in ``Mass-Transfer Induced Activity
   in Galaxies,'' I. Shlosman, ed. (Cambridge Univ. Press), p.~131
\reference{} Knapen, J.H., Beckman, J.E., Heller, C.H., Shlosman, I. \&
   de Jong, R.S. 1995, ApJ, 454, 623
\reference{} Mailino, R., Ruiz, M. \& Rieke, G.H. 1995, ApJ 446, 561
\reference{} Malkan, M.A., Gorjian, V. \& Tam, R., 1998, ApJS, 117, 25
\reference{} Martin, P., 1995, AJ, 109, 2428
\reference{} McLeod, K.K. \& Rieke, G.H. 1995, ApJ, 441, 96
\reference{} Moles, M., M\'arquez, I. \& P\'erez, E. 1995, ApJ, 438, 604
\reference{} Mulchaey, J. \& Regan, M., 1997, ApJ, 482, L135
\reference{} Mulchaey, J., Regan, M. \& Kundu, A., 1997, ApJS, 110, 299
\reference{} Peletier, R.F., Davies, R.L., Illingworth, G., Davis, L. \& 
   Cawson, M. 1990, AJ, 100, 1091
\reference{} Peletier, R.F., Knapen, J.H., Shlosman, I.,
   P\'erez-Ramirez, D., Nadeau, D., Doyon, R., Rodriguez-Espinosa, J.M. \&
   P\'erez-Garc\'\i a, A.M. 1999, ApJS, in press ({\bf Paper~I})
\reference{} Phinney, E.S.P. 1994, in ``Mass-Transfer Induced Activity in   
   Galaxies,'' I. Shlosman, ed. (Cambridge Univ. Press), p.~1
\reference{} Regan, M.W. \& Mulchaey, J.S. 1999, ApJ, in press   
\reference{} Sandage, A. \& Tammann, G. A. 1981, A Revised Shapley-Ames
Catalog of Bright Galaxies (Publ. 635; 1st ed.; Washington: Carnegie
Institution of Washington)
\reference{} Sellwood, J.A. \& Wilkinson, A. 1993, Rep. Prog. Phys., 56, 173
\reference{} Shaw, M.A., Combes, F., Axon, D.J. \& Wright, G.S. 1993, A\&A, 
   273, 31
\reference{} Shlosman, I. 1996, in Proc. Nobel Symp. on ``Barred Galaxies
   \& Circumnuclear Activity'', A. Sandqvist \& P.O. Lindblad, eds. 
   (Springer-Verlag), p.~141
\reference{} Shlosman, I. 1992, in ``Relationships Between Active Galactic  
   Nuclei \& Starburst Galaxies,'' A.V. Filippenko, ed. (ASP Conf. Series),
   p.~335
\reference{} Shlosman, I., Begelman, M.C. \& Frank, J. 1990, Nature,
   345, 679
\reference{} Shlosman, I., Frank, J. \& Begelman, M.C. 1989, Nature, 338, 45 
\reference{} Shlosman, I., Peletier, R.F. \& Knapen, J.H., 1999, 
   in preparation ({\bf Paper~III}) 
\reference{} Simkin, S.M., Su, H.J. \& Schwarz, M.P. 1980, ApJ, 237, 404
\reference{} Spillar, E.J., Oh, S.P., Johnson, P.E. \& Wenz, M. 1992, AJ 103,
   793
\reference{} Thronson, H. Jr., Hereld, M., Majewski, S., Greenhouse, M.,
   Johnson, P., Spillar, E., Woodward, C.E., Harper, D.A. \& Rauscher, B.J. 
   1989, ApJ, 343, 158
\end{references}
\end{document}